\documentclass[prb,twocolumn,floatfix,showpacs,amssymb]{revtex4}
\usepackage{graphicx}
\usepackage{dcolumn}
\usepackage{bm}

\begin{document}

\title{Sign of the crossed conductances
at a FSF double interface}
\author{R. M\'elin\footnote{melin@grenoble.cnrs.fr}}
\affiliation{
Centre de Recherches sur les Tr\`es Basses
Temp\'eratures (CRTBT\footnote{U.P.R. 5001 du CNRS, Laboratoire
conventionn\'e avec l'Universit\'e Joseph Fourier}),\\ CNRS, BP 166,
38042 Grenoble Cedex 9, France}
\author{D. Feinberg}
\affiliation{
Laboratoire d'Etudes des Propri\'et\'es Electroniques des
Solides (LEPES\footnote{U.P.R. 11 du CNRS, Laboratoire
conventionn\'e avec l'Universit\'e Joseph Fourier}),\\ CNRS, BP 166,
38042 Grenoble Cedex 9, France}

\begin{abstract}
Crossed conductance in hybrid Ferromagnet~/ Superconductor~/
Ferromagnet (FSF) structures results from the competition
between normal transmission and
Andreev reflection channels. Crossed Andreev reflection (CAR) and
elastic cotunneling (EC) 
between the ferromagnets
are dressed by local Andreev reflections, which play an important
role for transparent enough interfaces and intermediate
spin polarizations. This modifies the simple result previously obtained
at lowest order, and
can explain the sign of the crossed resistances in a
recent experiment [D. Beckmann {\sl et al.},
cond-mat/0404360].
This holds both
in the multiterminal hybrid structure model (where phase averaging
over the Fermi oscillations
is introduced ``by hand'' within the approximation of a single
non local process) and for
infinite planar interfaces (where phase averaging naturally results
in the microscopic solution with multiple 
non local processes).
\end{abstract}


\maketitle

\section{Introduction}

Andreev reflection~\cite{Andreev,Saint-James} is the mechanism by which
charge is transported at normal metal~/ superconductor (NS)
interfaces at voltages below
the superconducting gap. A spin-up electron from the N electrode
is reflected as a spin-down hole while a Cooper pair is
transfered in the superconductor.
Multiterminal geometries are of particular interest.
For instance the sign of the crossed conductance between
two electrodes connected to a same superconductor can be either
positive or negative depending on the relative strengths of the "normal" (electron-electron) and
"anomalous" (electron-hole) channels \cite{Lambert1,Lambert2,Jedema}.
Multiprobe structures are especially interesting
when the electrodes are ferromagnets. Actually, Andreev reflection also exists
in ferromagnet~/ superconductor (FS)
junctions but is suppressed as the spin polarization of the
ferromagnet increases~\cite{deJong,Soulen,Upa}.
It was shown\cite{Deutscher,Falci,Melin-JPCM,Melin-Feinberg-CAR,
Melin-Peysson,Chte,AB,Feinberg-des,Buttiker,Pistol,Lambert3,Beckmann}
that "crossed" Andreev reflections (CAR) between two ferromagnetic electrodes, where
spin-up electron and outgoing spin-down hole belong
to different electrodes, give rise to interesting magnetoresistance effects. For instance,
the latter process is favored by
antiparallel spin polarizations. The fundamental problem of calculating the
scattering matrix $S_{ij}$ of a NSN structure has been solved in some limits. The authors of
 Ref. \onlinecite{Lambert3} use a "fork" geometry where the two electrodes merge at the
superconductor. In contrast, in Refs. \onlinecite{Deutscher,Falci,Melin-JPCM} (like in
a recent experiment \cite{Beckmann}),
two distinct F/S contacts (denoted here as $a,b$) are separated by a distance $R$.
Yamashita et al. \cite{yama}
used a generalization of the BTK approach
in this geometry\cite{BTK}.
As a common result, CAR is possible if $R<\xi$,
the superconducting coherence length, but is also reduced by an
algebraic factor. Andreev reflection becomes in this geometry a
genuine mesoscopic effect, non-local at the scale of the
physical contacts. On the other hand,
in the normal channel, transmission between the two electrodes
is also possible. Since it is
spin-conserving, it is favored by parallel spin polarizations. In a tunnel model for the
two F/S junctions, CAR and EC involve the virtual creation and propagation of a
quasiparticle in the superconductor. For the normal channel, this is similar to
cotunneling, introduced in the context of transmission across a Coulomb
blockaded quantum dot
\cite{cotun}.
The superconducting gap replacing the Coulomb energy,
and the normal crossed process
has been  called "elastic cotunneling" in Ref.~\onlinecite{Falci}.

Other manifestations of spatially separated pair correlations
were obtained in the study of equilibrium properties of FSF
trilayers\cite{Melin-JPCM,Apinyan,Jirari,Buzdin-Daumens,MF-tri,Melin-long}.
It was shown\cite{Melin-JPCM,Apinyan,Jirari}
within a model of multiterminal hybrid structure
that the self-consistent superconducting gap can be
larger in the parallel alignment. The same result was obtained for the
FSF trilayer with atomic thickness, for half-metal
ferromagnets\cite{Buzdin-Daumens} and Stoner ferromagnets\cite{MF-tri}.
However simulations with a finite thickness\cite{MF-tri,Melin-long}
showed that pair-breaking dominates for strong ferromagnets
as the thickness of the
superconductor is larger than the Fermi wave-length (with therefore
the superconducting gap larger in the antiparallel alignment).

The goal of the theory of crossed conductances at a FSF double
interface
is to calculate the conductances
of a FSF structure for any relative alignments of the spin polarizations.
In a first approximation we neglect the dependence of the
superconducting gap on the relative spin orientation in the
ferromagnets, which is a consistent assumption if the
size of the contacts is much smaller than the superconducting
coherence length \cite{thesis-blonder}.
The first approach to this problem was
through Landauer formalism\cite{Deutscher}.
Lowest order perturbation theory\cite{Falci} gives an interpretation
in terms of CAR and EC processes. The effect
of non collinear ferromagnets was also investigated\cite{Melin-Peysson,Chte},
with the aim of describing
transport of Cooper pairs at the interface between
a superconductor and a ferromagnet containing a domain wall.
The Josephson effect between two superconductors connected
by two spatially separated conduction channels was
also examined\cite{Melin-Peysson} and it was found that
there is no Josephson effect
within lowest order perturbation theory
unless the length of the ferromagnets
is smaller than the elastic mean free path, a condition that
is not usually verified in experiments.
Toy models for more complicated geometries
involving Aharonov-Bohm effects related to crossed correlations
were also investigated\cite{AB}.
Disorder effects were also discussed recently for tunnel
interfaces\cite{Feinberg-des,Chte}
and it was found that the geometrical reduction is less severe
in the presence of disorder in the superconductor (dirty limit).
Another geometry with a normal metal island connected to
one superconductor and two ferromagnets was proposed in
Ref.~\onlinecite{Buttiker}.
Noise correlations were
also discussed~\cite{Tadei} and
recently within lowest order perturbation
theory\cite{Pistol}.

CAR and EC were probed
in a recent experiment by Beckmann {\it et al.}\cite{Beckmann}.
Driving a current
through one contact induces a voltage in the other one.
The overall experimental results (the non-local resistance) are
in agreement with theory
except for the sign of the effect: it was
predicted theoretically\cite{Lambert3,Falci}
that for tunnel barriers the induced currents in the parallel and
antiparallel spin orientations
have an opposite sign whereas in
experiments\cite{Beckmann} they have the same sign.
Here we resolve this apparent contradiction
by noting that the large interface transparencies used in
experiments imply that the CAR and EC processes are
``dressed'' by local Andreev reflections at the two FS
interfaces. This dressing is easily transcripted at zero temperature
in terms of a perturbative
expansion for the Keldysh Green's functions. Due to the strong
damping of quasiparticles
in the superconductor,
a reasonable approximation involves a single non local propagator while
local ones are treated to all orders. We also carry out
numerical simulations of infinite planar interfaces in which
multiple non local processes are taken into account and find a good
agreement with the analytical approach.

The article is organized as follows.
Preliminaries are given in section~\ref{sec:prelim}.
The properties of multiterminal hybrid structures are investigated
in section~\ref{sec:multiter}. Infinite planar interfaces are
investigated in section~\ref{sec:infplanar}. Concluding remarks
are given in section~\ref{sec:conclu}.

\section{Preliminaries}
\label{sec:prelim}
In this section we provide the form of the Green's functions
and Hamiltonians that we use throughout the article.

\subsection{Hamiltonians}
The superconductor is described by the BCS Hamiltonian\cite{Tinkham}:
\begin{eqnarray}
\label{eq:H-BCS}
{\cal H}_{\rm BCS}&=&\sum_{{\bf k},\sigma}
\epsilon(k) c_{{\bf k},\sigma}^+ c_{{\bf k},\sigma}\\\nonumber
&+& \Delta \sum_{\bf k} \left(
c_{{\bf k},\uparrow}^+ c_{{\bf k},\downarrow}^+
+ c_{{\bf k},\downarrow} c_{{\bf k},\uparrow}
\right)
,
\end{eqnarray}
where $\epsilon(k)=\hbar^2 k^2 / 2 m$ is the free electron
dispersion relation
and $\Delta$ the superconducting gap.
The ferromagnets are described by the Stoner model
\begin{eqnarray}
{\cal H}_{\rm Stoner}&=&\sum_{{\bf k},\sigma}
\epsilon(k) c_{{\bf k},\sigma}^+ c_{{\bf k},\sigma}\\\nonumber
&-& h_{\rm ex} \sum_{\bf k}
\left(
c_{{\bf k},\uparrow}^+ c_{{\bf k},\uparrow}
- c_{{\bf k},\downarrow}^+ c_{{\bf k},\downarrow}
\right)
,
\end{eqnarray}
where $h_{\rm ex}$ is the exchange field.

\subsection{Green's functions in reciprocal space}
For collinear magnetizations
the Nambu representation corresponds to two $2 \times 2$ matrices,
one in the sector $S_z=1/2$ and the other in the sector $S_z=-1/2$
($S_z$ is the projection of the
spin along the quantization axis chosen parallel to the exchange field).
The advanced
Green's function in the sector $S_z=1/2$ is given at
zero temperature by
\begin{eqnarray}
\label{eq:Green-def}
&& \hat{g}_{{\bf x},{\bf y}}^A(t,t') = -i \theta(t-t')\\
\nonumber
&& \left( \begin{array}{cc}
\langle \{ c_{{\bf x},\uparrow}(t) , c_{{\bf y},\uparrow}^+(t') \} \rangle &
\langle \{ c_{{\bf x},\uparrow}(t) , c_{{\bf y},\downarrow}(t') \} \rangle \\
\langle \{ c_{{\bf x},\downarrow}^+(t) , c_{{\bf y},\uparrow}^+(t')\} \rangle &
\langle \{ c_{{\bf x},\downarrow}^+(t) , c_{{\bf y},\downarrow}(t') \} \rangle
\end{array} \right)
,
\end{eqnarray}
where ${\bf x}$ and ${\bf y}$ are two arbitrary sites
and $\{.,.\}$ is an anticommutator.
Using the Hamiltonian (\ref{eq:H-BCS}) one obtains the advanced Green's function
in reciprocal space\cite{Abrikosov}:
\begin{eqnarray}
\label{eq:galpha11}
\label{eq:gS-11}
g^{1,1,A}(\xi,\omega)&=&
\frac{u_k^2}{(\omega-\mu_S)-E_k-i\eta_S}\\
\nonumber
&+&
\frac{v_k^2}{(\omega-\mu_S)+E_k-i\eta_S}\\
\label{eq:falpha12}
\label{eq:gS-12}
f^{1,2,A}(\xi,\omega)&=&
-\Delta\frac{1}{(\omega-\mu_S)-E_k-i\eta_S}\\
\nonumber
&&\times\frac{1}{(\omega-\mu_S)+E_k-i\eta_S}
,
\end{eqnarray}
where $\mu_S$ is the chemical potential in the superconductor,
$\xi=\hbar^2 k^2 / 2 m - \mu_S$ is the kinetic
energy with respect to the Fermi level,
$E_k=\sqrt{\Delta^2+\xi_k^2}$ is the quasiparticle energy,
$u_k^2=(1+\xi_k/E)/2$ and
$v_k^2=(1-\xi_k/E)/2$ are the BCS coherence factors,
and $\eta_S$ is a small parameter related to inelastic processes in the
superconductor (a typical\cite{Kaplan,Cuevas} value in experiments
is $\eta_S=10^{-2} \Delta$).

The ``11'' component of the Green's function of a ferromagnet is given by
\begin{equation}
\label{eq:gF}
g_{a,a}^{1,1}(\xi,\omega)
= \frac{1}{(\omega-\mu_a)-\xi+h_{\rm ex} -i \eta_F}
,
\end{equation}
where $\mu_a$ is the chemical potential in the ferromagnet ``a''.
A similar expression is obtained for $g_{a,a}^{2,2}$.
The parameter $\eta_F$ contains information about decoherence
in the ferromagnet\cite{sin2phi}.

\subsection{Green's functions in real space}
\begin{figure}
\includegraphics [width=1. \linewidth]{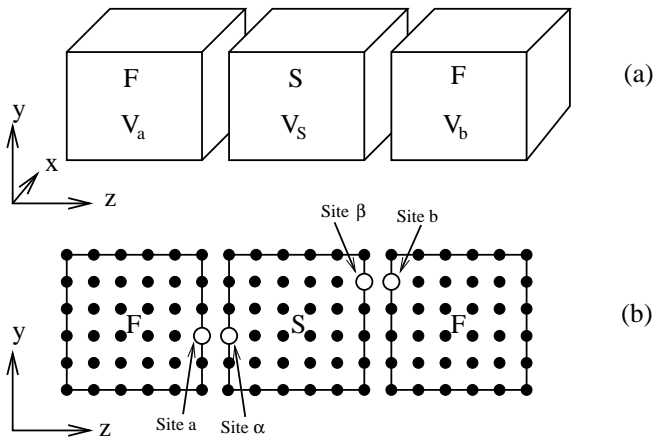}
\caption{(a) Schematic representation of the Andreev reflection
geometry in a FSF structure. (b) Schematic representation of the
tight-binding Hamiltonian cubic lattice and notation for the sites
$a$, $\alpha$, $\beta$ and $b$. (b) corresponds to a cut in the
$(y,z)$ plane.
Voltages $V_a$ and $V_b$ are
applied on the two ferromagnets and a voltage $V_S$ is applied on
the superconductor.
\label{fig:schema}
}
\end{figure}
It is useful in transport theory to manipulate Green's functions
parameterized by spatial variables. To carry out the Fourier
transform we define a short-scale cut-off
corresponding to the lattice spacing $a_0$
of a cubic lattice tight-binding model
where electrons hop between neighboring
discrete sites like in Ref.~\onlinecite{CCNSJ}
(see Fig.~\ref{fig:schema}).
The tight-binding and continuous models are equivalent since
we consider low energies compared to the band-width.

The local Green's functions of a superconductor is defined as
\begin{equation}
\hat{g}_{\alpha,\alpha}(\omega) = \int
\frac{d^3 {\bf k}}{(2 \pi)^3}
\hat{g}(\xi,\omega)
,
\end{equation}
where $\alpha$ is a site in the superconductor (see Fig.~\ref{fig:schema}),
and
where $\hat{g}(\xi,\omega)$ is given by Eqs.~(\ref{eq:gS-11})
and~(\ref{eq:gS-12}).
Evaluating the integral over wave vector
leads to
\begin{eqnarray}
&&\hat{g}_{\alpha,\alpha}(\omega)=\hat{g}_{\beta,\beta}(\omega)\\
\nonumber
&=& \frac{\pi \rho_S}{\sqrt{\Delta^2-(\omega-\mu_S)^2}}
\left[ \begin{array}{cc} -(\omega-\mu_S) & \Delta \\
\Delta & - (\omega-\mu_S) \end{array} \right]
,
\end{eqnarray}
where $\rho_S$ is the normal state density of state,
$\omega$ is the energy and $\Delta$ is
the superconducting gap. The Green's functions
of the superconductor take the form \cite{error}
\begin{eqnarray}
\nonumber
&&\hat{g}_{\alpha,\beta}(\omega)=
\frac{\pi \rho_S}{k_F R} \exp{\left(-\frac{R}{\xi(\omega)}\right)}\\
\nonumber
&&\left\{\frac{\sin{(k_F R)}}{\sqrt{\Delta^2-(\omega-\mu_S)^2}}
\left[ \begin{array}{cc} -(\omega-\mu_S) & \Delta \\
\Delta & -(\omega-\mu_S)
\end{array} \right] \right.\\
&+& \left. \cos{(k_F R)}
\left[ \begin{array}{cc} -1 & 0 \\ 0 & 1 \end{array} \right]
\right\}
,
\label{eq:g-nonloc-S}
\label{eq:GNL}
\end{eqnarray}
where $\alpha$ and $\beta$ are two sites in the superconductor
(see Fig.~\ref{fig:schema}).
$\xi(\omega)=\hbar v_F/\sqrt{\Delta^2-(\omega-\mu_S)^2}$ is the
BCS coherence length.
The local Green's functions of the ferromagnet is given by
\begin{equation}
\label{eq:g-loc-ferro}
\hat{g}_{a,a} = i \pi \left[ \begin{array}{cc}
\rho_{a,\uparrow} & 0 \\ 0 & \rho_{a,\downarrow}
\end{array} \right]
,
\end{equation}
where $\rho_{a,\uparrow}$ and $\rho_{a,\downarrow}$ correspond
to the spin-up and spin-down density of states in the ferromagnet
``a''. A spin polarization $P$ corresponds to
$\rho_{a,\uparrow}=\rho_F (1+P_a)$ and
$\rho_{a,\downarrow}=\rho_F (1-P_a)$. We discard the energy dependence
in (\ref{eq:g-loc-ferro}) since we are interested in energies
of order $\Delta$, much
smaller than the exchange field $h_{\rm ex}$ related to the spin
polarization by $P\simeq h_{\rm ex}/\epsilon_F$, for
$h_{\rm ex}$ small compared to $\epsilon_F$.

\subsection{Mixed Green's functions}
\label{sec:mixed}
To describe infinite planar contacts we use Green's functions
parameterized by the distance $R$ along the $z$ axis and the
two wave-vectors $k_x$ and $k_y$.
This parameterization is well suited for a situation where translation
invariance holds in the $(x,y)$ plane parallel to the interface but not in the
$z$ direction perpendicular to the interface.
The kinetic energy $\xi$ is separated into the sum of
$\xi_\parallel=\hbar^2 (k_x^2+k_y^2)/2m-\mu_S$ and
$\xi_\perp=\hbar^2 k_z^2 / 2m$. The mixed Green's function is
defined by
\begin{equation}
\hat{g}(R,\xi_\parallel,\omega)=
\int \frac{d k_z}{2 \pi} e^{i k_z R}
{\hat g}(\xi,\omega)
,
\end{equation}
where ${\hat g}(\xi,\omega)$ is defined by Eqs.~(\ref{eq:gS-11})
and~(\ref{eq:gS-12}).
After evaluating the Fourier transform with
respect to $k_z$ we obtain
\begin{eqnarray}
\label{eq:g-S-ip}
&& \hat{g}^A(R,\xi_\parallel,\omega) =
\frac{\sqrt{2 m a_0^2}}{2 \hbar} \exp{(i k_z R)}\\\nonumber
&&\times \frac{1}{\sqrt{ i
\sqrt{\Delta^2-(\omega-\mu_S-i \eta_S)^2}
-\xi_\parallel}} \\\nonumber
&&\left\{ \frac{1}
{\sqrt{\Delta^2-(\omega-\mu_S-i\eta_S)^2}}\right.\\\nonumber
&&\times\left.
\left[ \begin{array}{cc}
\omega-\mu_S-i\eta_S & -\Delta \\ -\Delta &
\omega-\mu_S-i\eta_S \end{array}
\right]
+ i \left[ \begin{array}{cc}
1 & 0 \\ 0 & -1 \end{array} \right] \right\}
,
\end{eqnarray}
with
\begin{equation}
k_z = \frac{\sqrt{2 m}}{\hbar} \sqrt{ i
\sqrt{(\omega-\mu_S-i \eta_S)^2-\Delta^2}
-\xi_\parallel}
.
\end{equation}

The mixed Green's function of a ferromagnet is given by
\begin{eqnarray}
\label{eq:g-F-ip}
g_{1,1}^A(R,\xi_\parallel,\omega) &=&
-\frac{\sqrt{2 m a_0^2}}{\hbar} \exp{(i k_z R)}\\\nonumber
&&\times \frac{1}{\sqrt{-(\omega-\mu_a)+i \eta_F
+\xi_\parallel-h}}
,
\end{eqnarray}
with
\begin{equation}
k_z = \frac{\sqrt{2 m}}{\hbar}
\sqrt{\omega-\mu_a-i \eta_F + h_{\rm ex}}
.
\end{equation}

\subsection{Transport properties}

The Green's functions $\hat{G}_{i,j}$ of the connected system are
obtained by solving the Dyson equation. In a compact notation the
Dyson equation takes the form $\hat{G}=\hat{g}+\hat{g} \otimes
\hat{\Sigma} \otimes \hat{G}$, where $\hat\Sigma$ is the self-energy
corresponding to the tunnel Hamiltonian, $\otimes$ is a summation
over spatial variables and a convolution over time variables,
and $\hat{g}$ is the Green's function of the disconnected
system with $\hat{\Sigma}=0$.

Tranport properties\cite{Cuevas,CCNSJ}
are obtained by evaluating the Keldysh
Green's function
\begin{equation}
\hat{G}^{+,-}
= [\hat{I}+\hat{G}^R \otimes \hat{\Sigma}]
\otimes \hat{g}^{+,-} \otimes
[\hat{I}+ \hat{\Sigma} \otimes \hat{G}^A ]
.
\end{equation}
The spin-up current through
the link $a$-$\alpha$ is given by
\begin{eqnarray}
\nonumber
I_{a,\alpha}&=&\frac{e}{2 h} \int d \omega
\mbox{Tr}
\left\{ \left[\hat{t}_{a,\alpha} \hat{G}^{+,-}_{\alpha,a}(\omega)
- \hat{t}_{\alpha,a} \hat{G}^{+,-}_{a,\alpha}(\omega) \right]
\hat{\sigma}^z \right\}\\
&+& (h_{\rm ex} \rightarrow - h_{\rm ex})
,
\label{eq:I-a-al}
\end{eqnarray}
where the trace is a summation over the ``11'' and ``22'' components
of the Green's function in the Nambu representation,
and $\hat{\sigma}^z$ is one of the Pauli matrices.
The term $(h_{\rm ex} \rightarrow - h_{\rm ex})$ corresponds
to a summation over the ``33'' and ``44'' components in the
$4 \times 4$ Nambu representation\cite{Melin-Peysson}
in the sector
$S_z=-1/2$. For a normal metal the four components of the
current are equal and we recover the usual form of the transport
formula\cite{Cuevas,CCNSJ}.

\section{Multiterminal hybrid structures}
\label{sec:multiter}
\subsection{Transport formula}

We start with the geometry on
Fig.~\ref{fig:schema}. In the
Hamiltonian approach, the exact geometry is unimportant : the results
are the same for
two lateral contacts (like in the experiment \cite{Beckmann}) or
opposite contacts (Fig.~\ref{fig:schema}).
Moreover we suppose in this section that the distance between the
contacts is large compared to their transverse dimension so that
the distance between sites $\alpha$ and $\beta$
(see Fig.~\ref{fig:schema}) can be
considered as approximately independent on the choice of
the sites $\alpha$ and $\beta$ at the interfaces. This hypothesis
does not hold in the infinite planar geometry (see
section~\ref{sec:infinite}).
The total current flowing between the
two electrodes is the sum of two
contributions : i) the CAR current (Andreev process) circulating from
both electrodes "a"
and "b", and ii) the EC current flowing from "a" to "b". Those are given by
\begin{eqnarray}
\label{eq:I-CAR-multi}
&&I_{\rm CAR}(V_a,V_b) = 2 \pi^2 t_a^2 t_b^2 \frac{e}{h}
\rho_{a,\uparrow} \rho_{b,\downarrow} \\\nonumber
&& \times \left\{ \left[\int_{-e V_b}^{e V_a}
+ \int_{-e V_a}^{e V_b}\right]d\omega
\overline{G_{\alpha,\beta}^{1,2,A}(\omega)
G_{\beta,\alpha}^{2,1,R}(\omega)}
\right\}\\\nonumber
&&+ 2 \pi^2 t_a^2 t_b^2 \frac{e}{h}
\rho_{a,\downarrow} \rho_{b,\uparrow} \\\nonumber
&& \times \left\{ \left[\int_{-e V_b}^{e V_a}
+ \int_{-e V_a}^{e V_b}\right]d\omega
\overline{G_{\alpha,\beta}^{2,1,A}(\omega)
G_{\beta,\alpha}^{1,2,R}(\omega)}
\right\}
,
\end{eqnarray}
and
\begin{eqnarray}
\label{eq:I-EC-multi}
&& I_{\rm EC}(V_a,V_b) = 2 \pi^2 t_a^2 t_b^2 \frac{e}{h}
\rho_{a,\uparrow} \rho_{b,\uparrow} \\\nonumber
&& \times \left\{ \left[ \int_{e V_b}^{e V_a}
+ \int_{-e V_a}^{-e V_b} \right] d\omega
\overline{G_{\alpha,\beta}^{1,1,A}(\omega)
G_{\beta,\alpha}^{1,1,R}(\omega)} \right.\\\nonumber
&&+ 2 \pi^2 t_a^2 t_b^2 \frac{e}{h}
\rho_{a,\downarrow} \rho_{b,\downarrow} \\ \nonumber
&&\times \left\{ \left[\int_{e V_b}^{e V_a}
+ \int_{-e V_a}^{-e V_b} \right] d\omega
\overline{G_{\alpha,\beta}^{2,2,A}(\omega)
G_{\beta,\alpha}^{2,2,R}(\omega)}
\right\}
.
\end{eqnarray}
Eq.~(\ref{eq:I-a-al}) is valid for all values of the voltages
$V_a$, $V_b$ and $V_S$. Eqs.~(\ref{eq:I-CAR-multi}) and
(\ref{eq:I-EC-multi}) are valid for all values of $V_a$ and $V_b$
and we supposed that $V_S=0$
since one reference voltage can be chosen equal
to zero.

The crossed conductance is defined \cite{Lambert2} as
${\cal G}_{ab}(V_a,V_b)=\partial I_{a,\alpha}(V_a,V_b) /
\partial V_b$, where $I_{a,\alpha}(V_a,V_b)=
I_{\rm CAR}(V_a,V_b)+I_{\rm EC}(V_a,V_b)$ is the total crossed current
through the link $\alpha$-$a$.
In the following we derive the currents
(\ref{eq:I-CAR-multi}) and
(\ref{eq:I-EC-multi}) with respect to the voltage $V_b$. One
contribution to the derivative is due to the upper and lower
bounds in the integrals. A second contribution is due to the
dependence of the Green's functions on $V_b$. However we deduce
from Eq.~(\ref{eq:gF}) that $\hat{g}_{b,b}$ is independent on
$V_b$ because the chemical potential is subtracted both in
the energy $\omega-\mu_b$ and in the kinetic energy $\xi$.
This second contribution to the
derivative is thus vanishingly small.
The crossed conductance can be
expressed in terms of the
spin polarizations $P_a$ and $P_b$:
\begin{eqnarray}
\label{eq:F1}
{\cal G}_{ab}&=&
4\pi^2 t_a^2 t_b^2 \frac{e}{h}\rho_{Fa}\rho_{Fb}
\left\{ \right.\\
&&\left.(1-P_aP_b)
\overline{G_{\alpha,\beta}^{1,2,A}(e V_b)
G_{\beta,\alpha}^{2,1,R}(e V_b)}
\right. \nonumber \\ \nonumber
\label{eq:F2}
&&-\left. (1+P_aP_b)
\overline{G_{\alpha,\beta}^{1,1,A}(e V_b)
G_{\beta,\alpha}^{1,1,R}(e V_b)} \right.\\\nonumber
&& \left.
+ (e V_b \rightarrow -e V_b)\right\}
.
\end{eqnarray}
Averaging over the Fermi oscillations\cite{Falci,Melin-Feinberg-CAR}
is used to simulate an extended contact as the sum of channels
in parallel with a distribution of length $R$:
\begin{eqnarray}
\overline{G_{\alpha,\beta}^{1,2,A}(\omega)
G_{\beta,\alpha}^{2,1,R}(\omega)}
&=& \frac{1}{\Delta R}
\int_{R-\Delta R/2}^{R+\Delta R/2} d R\\\nonumber
&& \times G_{\alpha,\beta}^{1,2,A}(R,\omega)
G_{\beta,\alpha}^{2,1,R}(R,\omega)
,
\end{eqnarray}
with $\Delta R=\lambda_F=2 \pi/k_F$.

The tunneling limit involves bare Green's functions,
which satisfy
$\overline{g_{\alpha,\beta}^{1,2,A}(\omega)
g_{\beta,\alpha}^{2,1,R}(\omega)}=
\overline{g_{\alpha,\beta}^{1,1,A}(\omega)
g_{\beta,\alpha}^{1,1,R}(\omega)}$, therefore the crossed conductances
for parallel
($P_a=P_b$) and antiparallel ($P_a=-P_b$) spin polarizations are opposite.
As a result, the crossed conductance is zero in absence of spin polarization \cite{Falci}. Similarly,
the crossed conductance at a mixed FSN interface (where only one electrode is ferromagnetic)
is zero.   As shown below, these symmetry properties do not hold anymore
 for transparent interfaces, where dressing of the
quasiparticle propagators occurs in the superconductor.
\begin{figure}
\includegraphics [width=.9 \linewidth]{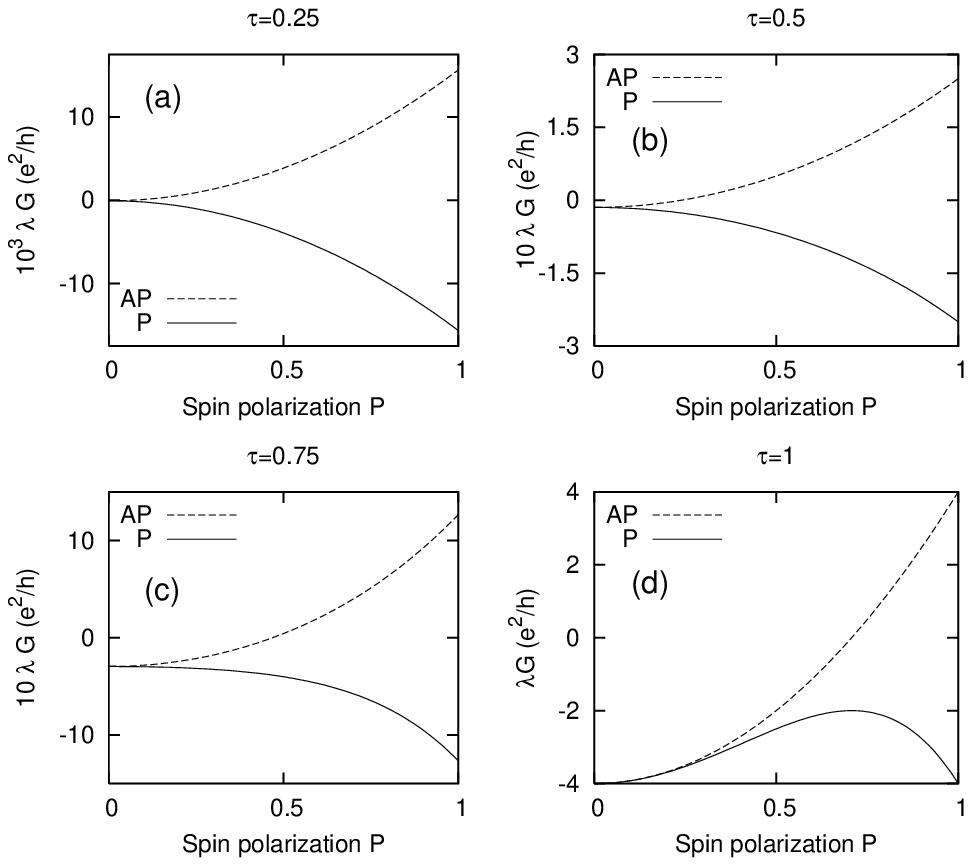}
\caption{Variation of the normalized linear crossed conductances
$\lambda {\cal G}_{\rm P}(P)$ (solid line) and
$\lambda {\cal G}_{\rm AP}(P)$ (dashed line)
as a function of the spin polarization~$P$,
for $\tau=0.25$ (a), $\tau=0.5$ (b),
$\tau=0.75$ (c), $\tau=2$ (d).
Note the different scaling factors on the
conductance axis.
and $\tau=1$
\label{fig:Fig1}
}
\end{figure}
\subsection{Perturbative expansion in $1/(k_F R)$}
Now we revisit the perturbative
expansion\cite{Falci,Melin-Feinberg-CAR,Melin-Peysson}
used to describe
crossed Andreev reflection.
We suppose that the contacts $a$-$\alpha$ and $b$-$\beta$ can be
highly transparent
so that multiple Andreev reflections take place locally. However
$k_F R$ is large so that in a first approximation
a single non local process
is included in $\hat{G}_{\alpha,\beta}$. This leads to
\begin{equation}
\label{eq:G-approx}
\hat{G}_{\alpha,\beta} = \hat{M}^{(0)}_{\alpha,\alpha}
\hat{g}_{\alpha,\beta} \hat{N}^{(0)}_{\beta,\beta}
,
\end{equation}
where $\hat{M}^{(0)}_{\alpha,\alpha}$
($\hat{N}^{(0)}_{\beta,\beta}$)
describes the dressing
by multiple Andreev reflections at the contact $a$-$\alpha$
($b$-$\beta$). $\hat{M}^{(0)}_{\alpha,\alpha}$ is determined by
the equation $\hat{G}^{(0)}_{\alpha,\alpha}=M^{(0)}_{\alpha,\alpha}
\hat{g}_{\alpha,\alpha}$, where $\hat{G}^{(0)}_{\alpha,\alpha}$ is
the fully dressed propagator at site $\alpha$ with $t_{b,\beta}=0$.
We find easily
\begin{eqnarray}
\hat{M}^{(0)}_{\alpha,\alpha}
&=&\left[ \hat{I} - \hat{g}_{\alpha,\alpha}
\hat{t}_{\alpha,a} \hat{g}_{a,a} \hat{t}_{a,\alpha} \right]^{-1}\\
\hat{N}^{(0)}_{\beta,\beta}
&=&\left[ \hat{I} - \hat{t}_{\beta,b} \hat{g}_{b,b}
\hat{t}_{b,\beta} \hat{g}_{\beta,\beta} \right]^{-1}
.
\end{eqnarray}
Generalizing Ref.~\onlinecite{Falci} we evaluate the phase
averaging of the following Green's functions:
\begin{eqnarray}
\label{eq:av1}
\overline{\left(g_{\alpha,\beta}^{1,1} \right)^2}
&=& \overline{\left(g_{\alpha,\beta}^{2,2} \right)^2}
= \overline{\left(g_{\alpha,\beta}^{1,2} \right)^2}\\
\nonumber
&=&\frac{\pi^2 \rho_S^2}{2 (k_F R)^2}
\exp{\left(-\frac{2 R}{\xi(\omega)}\right)}
\frac{\Delta^2}{\Delta^2-\omega^2}\\
\label{eq:av2}
\overline{g_{\alpha,\beta}^{1,1} g_{\alpha,\beta}^{2,2}}
&=&\frac{\pi^2 \rho_S^2}{2 (k_F R)^2}
\exp{\left(-\frac{2 R}{\xi(\omega)}\right)}
\frac{2 \omega^2-\Delta^2}{\Delta^2-\omega^2}\\
\label{eq:av3}
\overline{g_{\alpha,\beta}^{1,1} g_{\alpha,\beta}^{1,2}}
&=&\overline{g_{\alpha,\beta}^{2,2} g_{\alpha,\beta}^{1,2}}\\
\nonumber
&=&\frac{\pi^2 \rho_S^2}{2 (k_F R)^2}
\exp{\left(-\frac{2 R}{\xi(\omega)}\right)}
\frac{-\omega \Delta}{\Delta^2-\omega^2}
\end{eqnarray}
We use Eqs.~(\ref{eq:av1})-(\ref{eq:av3}) and
Eq.~(\ref{eq:G-approx})
to evaluate the
phase averaged Green's function in the CAR
and EC currents given by~(\ref{eq:I-CAR-multi})
and~(\ref{eq:I-EC-multi}).

\subsection{Crossed conductance in a FSF structure}
\begin{figure}
\includegraphics [width=.75 \linewidth]{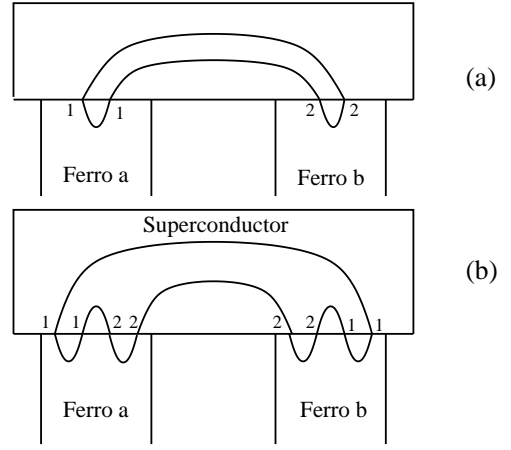}
\caption{(a) A CAR process to order
$\tau4$. (b) A CAR process to
order $\tau^8$ involving two local Andreev reflections.
The Nambu indices are indicated at each vertex. A similar diagram
can be drawn for EC process.
\label{fig:tau8}
}
\end{figure}

\subsubsection{Normalization of the tunnel amplitudes}

For a NN contact, highly transparent interfaces correspond
to $t=t_0$, where $t_0$ is such that
$\pi^2 t_0^2 \rho_N^2=1$. The NN contact conductance is
given by ${\cal G}_{NN}=(2 e^2 / h) \alpha$, with\cite{Cuevas}
\begin{equation}
\alpha= \frac{4 (t/t_0)^2}
{(1+ (t/t_0)^2)^2}
.
\end{equation}
The conductance of a NF contact is given by
${\cal G}_{NF}=(e^2 / h) (\alpha_\uparrow+\alpha_\downarrow)$, with
\begin{eqnarray}
\alpha_\uparrow &=&
\frac{4 \pi^2 t^2 \rho_N \rho_F (1+P)}
{ [ 1 + \pi^2 t^2 \rho_N \rho_F (1+P) ]^2}\\
\alpha_\downarrow &=&
\frac{4 \pi^2 t^2 \rho_N \rho_F (1-P)}
{ [ 1 + \pi^2 t^2 \rho_N \rho_F (1-P) ]^2}
\end{eqnarray}

The value $t_0$ of $t$ corresponding to
a perfect transmission in the spin-up channel is given by
\begin{equation}
\pi^2 t_0^2(P) \rho_N \rho_F (1+P)=1
\end{equation}
The conductance of a SF contact with
perfect transmission in the spin-up
channel ($t=t_0(P)$) is given by
\begin{equation}
{\cal G}_{NF}={e^2 \over h} \left(1 + \frac{4
{1-P \over 1+P}}{\left(1+{1-P\over 1+P} \right)^2} \right)
,
\end{equation}
varying between $e^2/h$ for half-metal ferromagnets
and $2 e^2/h$ in the absence of spin polarization.
In the following we normalize the hopping amplitude $t$
to the maximal value $t_0(1)$ of $t_0(P)$:
$t=\tau t_0(1)$, with $\tau$ between $0$ and $1$.

\subsubsection{Linear crossed conductance}

Taking $\omega \ll \Delta$, the linear crossed conductances corresponding to $V_b=0$
are given by
\begin{eqnarray}
\label{eq:GCAR-eps}
\lambda
{\cal G}_{\rm AP} &=& 4 \frac{e^2}{h} \frac{\tau^4}{
{\cal D}^2(P)} \left\{- \frac{\tau^4 (1-P^2)}
{{\cal D}^2(P)} + P^2 \right\}\\
\label{eq:GEC-eps}
\lambda
{\cal G}_{\rm P} &=& 4 \frac{e^2}{h} \frac{\tau^4}{
{\cal D}^2(P)} \\\nonumber
&& \left\{- \frac{\tau^4 (1-P^2)(1-2 P^2)}{
{\cal D}^2(P)} - P^2 \right\}
,
\end{eqnarray}
with ${\cal D}(P)=1+\tau^4(1-P^2)/4$,
$\lambda=2 (k_F R)^2 \exp{(2 R / \xi_0)}$ in the ballistic
limit, and $\lambda=2 (k_F l_e) (k_F R) \exp{(2 R / \xi_0)}$
in the diffusive limit\cite{Feinberg-des},
where $l_e$ is the elastic mean free path
and $\xi_0$ the coherence length at zero energy.
The variations of $\lambda {\cal G}_{\rm AP}(P)$ and
$\lambda {\cal G}_{\rm P}(P)$ are shown on Fig.~\ref{fig:Fig1}
for increasing values of $\tau$. First, one sees that at $P=0$ the crossed conductance
is negative, as if a normal metal would replace
the superconductor \cite{Lambert1}.
By contrast,
in the tunneling approach, an exact symmetry holds between CAR and EC, so that the
crossed conductance is zero for $P_a=0$ or $P_b=0$ \cite{Falci,Melin-Feinberg-CAR}.
The observed trend is more apparent for transparent contacts, and means
that EC processes dominate over CAR.
We obtain ${\cal G}_{\rm AP}(P) \simeq -{\cal G}_{\rm P}(P)$
only for small values of $\tau$
($\tau=0.25$ on Fig.~\ref{fig:Fig1}-(a)).
As $\tau$ is increased ${\cal G}_{\rm AP}(P)$ changes sign
as shown on Figs.~\ref{fig:Fig1}-(b), \ref{fig:Fig1}-(c)
and \ref{fig:Fig1}-(d).
For half-metal ferromagnets we have
${\cal G}_{\rm AP}(P)=-{\cal G}_{\rm P}(P)$ for all values
of $\tau$, as it can be seen from Eqs.~(\ref{eq:GCAR-eps})
and~(\ref{eq:GEC-eps}).
The key role in the sign changes
is played by processes of order $\tau^8$ like
the one represented on Fig.~\ref{fig:tau8}-(b).
Due to the opposite signs of
$\overline{(g_{\alpha,\beta}^{1,1})^2}$ and $\overline{g_{\alpha,\beta}^{1,1}
g_{\alpha,\beta}^{2,2}}$ for $\omega=0$
(see Eqs. (\ref{eq:av1}, \ref{eq:av2})), 
local Andreev dressing effects
tend to decrease $\overline{G_{\alpha,\beta}^{1,2,A}
G_{\beta,\alpha}^{2,1,R}}$ and increase $\overline{G_{\alpha,\beta}^{1,1,A}
G_{\beta,\alpha}^{1,1,R}}$, therefore CAR is weakened and EC
reinforced.

\subsubsection{Crossed conductance versus voltage}
\begin{figure}
\includegraphics [width=1 \linewidth]{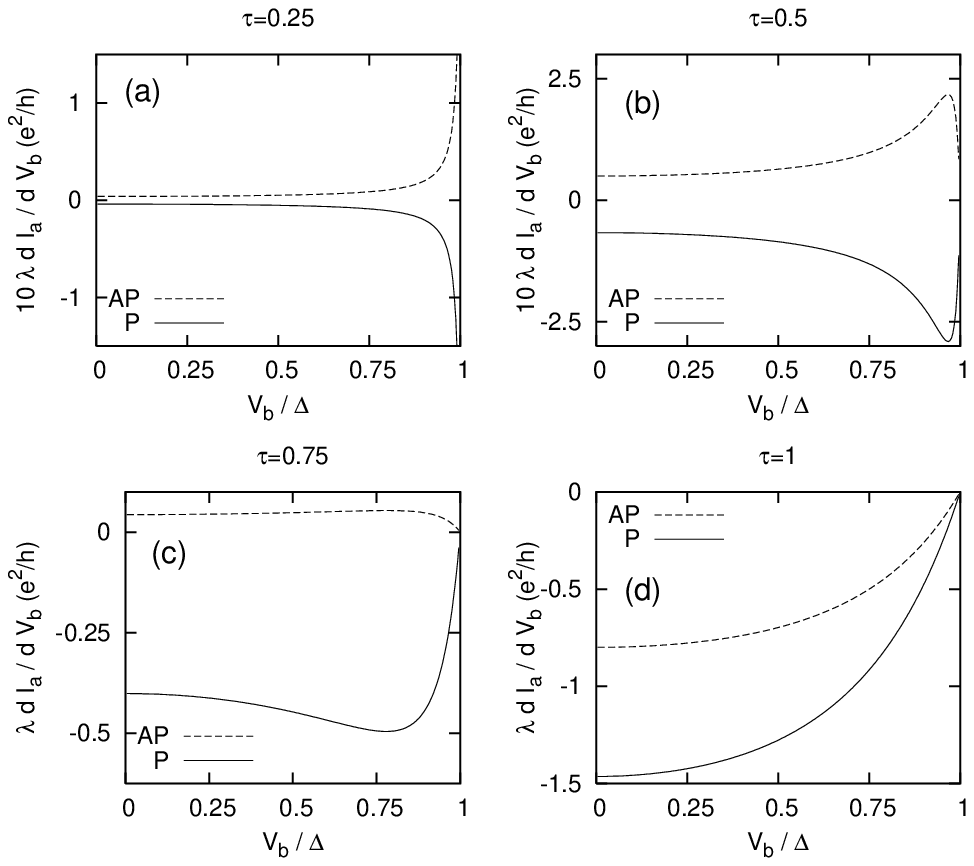}
\caption{Variation of the normalized crossed conductance
$\lambda(V_b) {\cal G}(V_b)=
\lambda \partial I_a / \partial V_b (V_a=0,V_b)$
as a function of the normalized voltage $V_b/\Delta$,
for $P=0.5$ and increasing values of
the dimensionless tunnel amplitude~$\tau$. The
parallel (antiparallel) alignment corresponds to the
solid (dashed) line.
\label{fig:Fig3}
}
\end{figure}
The variations of the crossed conductances as a function of
the voltage $V_b$ applied on electrode ``b'' are shown on
Fig.~\ref{fig:Fig3} for $P=0.5$.
For small interface transparencies ($\tau=0.25$ and $\tau=0.5$)
the crossed conductances in the parallel and
antiparallel alignments are approximately opposite in
the entire voltage range. The crossed conductance at
$V_b=\Delta$ is vanishingly small, both in the
parallel and antiparallel alignments and for arbitrary spin
polarizations.

For half-metal ferromagnets the crossed conductances take the form
\begin{eqnarray}
\label{eq:half}
\lambda(V_b)
{\cal G}_{\rm AP}(V_b)&=&
-\lambda(V_b){\cal G}_{\rm P}(V_b)=\\\nonumber
&&4 \tau^4 \frac{\Delta^2}{\Delta^2-V_b^2}
\left(1+ \tau^4 \frac{V_b^2}{\Delta^2-V_b^2}
\right)^{-2}
,
\end{eqnarray}
where $\lambda(V_b)$ has the same expression as $\lambda$
except that $\xi_0$ is replaced by the coherence length at
a finite energy $\xi_0 \Delta/\sqrt{\Delta^2-V_b^2}$.
Eq.~(\ref{eq:half}) is approximately equal to
$(8/\tau^4)(1-V_b/\Delta)$
if $\Delta-\omega \ll \tau^4 \Delta$,
and approximately equal to $4 \tau^4 \Delta^2/(\Delta^2-
V_b^2)$ if $\Delta-V_b \gg \tau^4 \Delta$.
For $\tau$ small
there is thus a maximum in the crossed
conductance\cite{Melin-Feinberg-CAR}
at a voltage $V_b \simeq \Delta (1-\tau^4/2)$.
This argument for $P=1$ is compatible with the behavior
of the crossed conductances for $V_b \alt \Delta$
and $P=0.5$ on Fig.~\ref{fig:Fig3}.

\subsubsection{Crossed conductance of a FSN or NSF junction}
The linear crossed conductance of a FSN (NSF) junction
\begin{equation}
\lambda {\cal G}=-\frac{e^2}{h}
\frac{8 \tau^8 (1-P^2)}
{(1+\tau^4(1-P^2)/4)(1+\tau^4/4)}
\end{equation}
is proportional to $\tau^8$ and thus very small for
tunnel interfaces. However it can take measurable values
for highly transparent interfaces and with a weakly polarized
ferromagnet or even a normal metal.

\subsection{Interpretation of the dressing by local processes}
One of the terms contributing to $G_{\alpha,\beta}^{1,1,A}
G_{\beta,\alpha}^{1,1,A}$ in the parallel alignment is
\begin{equation}
\label{eq:exp}
M_{\alpha,\alpha}^{(0),1,2}
g_{\alpha,\beta}^{2,1}
N_{\beta,\beta}^{(0),1,1}
M_{\beta,\beta}^{(0),1,2}
g_{\beta,\alpha}^{2,1}
N_{\alpha,\alpha}^{(0),1,1}
,
\end{equation}
from what we deduce
one of the terms of order $t_a^4 t_b^4$ contributing
$t_a^2 t_b^2 \rho_{a,\uparrow}
\rho_{b,\uparrow} G_{\alpha,\beta}^{1,1,A}
G_{\beta,\alpha}^{1,1,A}$:
\begin{eqnarray}
t_{\alpha,a}^{1,1} \rho_{a,a}^{1,1}
t_{a,\alpha}^{1,1}
g_{\alpha,\alpha}^{1,2}
t_{\alpha,a}^{2,2} g_{a,a}^{2,2,A}
t_{a,\alpha}^{2,2} g_{\alpha,\beta}^{2,1}\\
\times t_{\beta,b}^{1,1} \rho_{b,b}^{1,1} t_{b,\beta}^{1,1}
g_{\beta,\beta}^{1,2} t_{\beta,b}^{2,2}
g_{b,b}^{2,2,R} t_{b,\beta}^{2,2}
g_{\beta,\alpha}^{2,1}
\end{eqnarray}
This process corresponds to a diagram in which a spin-up electron
from electrode ``a'' is locally
Andreev reflected as a spin-down hole that makes an excursion
in electrode ``a'', is converted in a spin-up electron in electrode
``b'' through a CAR. The spin-up electron
in electrode ``b'' undergoes a local Andreev reflection, is converted
in a spin-down hole that is transformed in a spin-up electron
by a CAR.
This process is not possible for half-metal ferromagnets, in which case
the linear conductance is equal to the bare conductance,
proportional to $\tau^4$.

\section{Infinite planar interfaces with a bulk superconductor}
\label{sec:infplanar}
\label{sec:infinite}

\begin{figure}
\includegraphics [width=1. \linewidth]{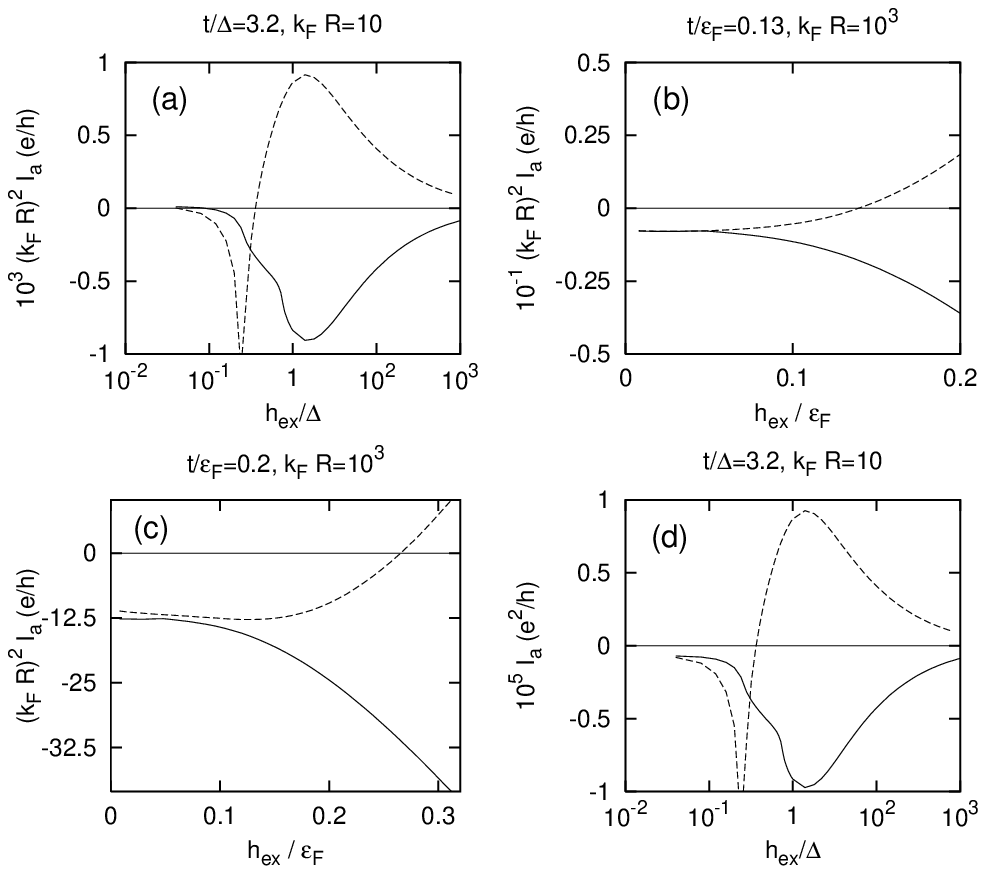}
\caption{Variation of the total current $I_a$ through electrode ``a''
evaluated with $e V_a=0$ and $e V_b= \Delta / 2$
as a function of the exchange field,
with $\Delta/\epsilon_F=4 \times 10^{-3}$.
The parallel (antiparallel) alignment corresponds to the
solid (dashed) line.
(a), (b) and (c) correspond to a full Green's function calculation
with CAR and EC to all orders.
(d) corresponds to the
approximation given by Eq.~(\ref{eq:G-approx}).
We use $\eta_S/\Delta=10^{-1}$ and $\eta_F/\epsilon_F=0.4$ but
obtained similar results for $\eta_F/\Delta=10^{-1}$.
\label{fig:Fig2}
}
\end{figure}

Now we consider infinite planar 2D interfaces connecting a bulk
3D superconductor to bulk 3D ferromagnets.
Within this approach we do not have to impose ``by hand''
phase averaging as we did in the previous section
(see Eqs.~(\ref{eq:av1})-(\ref{eq:av3})).
We use the mixed Green's functions introduced in section~\ref{sec:mixed}.
The evaluation of the integrals in the expression of the CAR
and EC currents is briefly described in Appendix~\ref{app:change}.
A realistic model would involve an estimate of the inverse proximity
effect, {\sl i.e.} how the superconducting gap depends on the
relative spin orientation of the ferromagnetic electrodes
\cite{Apinyan,Jirari,Buzdin-Daumens,MF-tri,Melin-long}.
Here we consider the infinite planar limit just as a test of the
approach in the
preceding section and do not impose self-consistency on the
superconducting gap.

We calculate the crossed
current through electrode ``a'' with a finite voltage
$V_b$ on electrode ``b'', and with $V_a=0$ and
suppose that the distance between the ferromagnets is
much larger than the Fermi wave-length.
The variations of the current evaluated at $V_b=\Delta/2$ as
a function of $h_{\rm ex}/\Delta$ is shown on Fig.~\ref{fig:Fig2}
for different values of the distance $R$ between the contacts.
For small values of $t/\Delta$ we obtain one Andreev bound state
for weak ferromagnets for which
$h_{\rm ex}$ and $\Delta$ have the same order
of magnitude (see Fig.~\ref{fig:Fig2}-(a)).
Andreev bound states for weak ferromagnets were
already discussed elsewhere\cite{Melin-long}.
For larger values of $t/\Delta$ and larger values of $R/a_0$
(see Fig.~\ref{fig:Fig2}-(b))
we obtain $I_{\rm AP} \simeq I_{\rm P}$ up to relatively large values of
$h_{\rm ex}/\Delta$ and $I_{\rm AP} \simeq - I_{\rm P}$ for the largest
values of $h_{\rm ex}/\Delta$. The overall variation on
Fig.~\ref{fig:Fig2}-(a), Fig.~\ref{fig:Fig2}-(b) and
Fig.~\ref{fig:Fig2}-(c)
looks like Fig.~\ref{fig:Fig1}-(a),
Fig.~\ref{fig:Fig1}-(b) and Fig.~\ref{fig:Fig1}-(c).
Fig.~\ref{fig:Fig2}-(d) corresponds to the same parameters
as Fig.~\ref{fig:Fig2}-(a) except that
we used the approximation~(\ref{eq:G-approx}) of a single
non local process
on Fig.~\ref{fig:Fig2}-(d)
whereas Fig.~\ref{fig:Fig2}-(a) was obtained without approximation.
The two variations are almost identical,
except for the regime $h_{\rm ex}\ll\Delta$. The agreement is even better
for larger values of $R$ (not shown on Fig.~\ref{fig:Fig2}). This
illustrates the validity of the approximation~(\ref{eq:G-approx})
in which we keep a single CAR.

\section{Conclusions}
\label{sec:conclu}

To conclude we have proposed a model that may constitute
an explanation to the signs of
the crossed currents in the recent experiment
by Beckmann {\it et al.}\cite{Beckmann}. 
In the experiment,
a current is injected through electrode $b$, and a voltage $V_a$ is
measured. Assuming that the local conductances ${\cal G}_{aa}$,
${\cal G}_{bb}$ are much larger than ${\cal G}_{ab}$, ${\cal G}_{ba}$,
one simply gets
the non-local resistance
${\cal R}_{ab}^{\rm P,AP}
\sim -{\cal G}_{ab}^{\rm P,AP}/{\cal G}_{aa}{\cal G}_{bb}$.
For small interface
transparencies we recover the results of
perturbation theory to order $t_a^2 t_b^2$ : the crossed
conductance in the antiparallel alignment ${\cal G}_{ab}^{\rm AP}$ is opposite to the
crossed conductance ${\cal G}_{ab}^{\rm P}$ in the parallel alignment, and
is zero if $P_a=0$ or $P_b=0$. For
larger interface transparencies, the propagators of crossed
Andreev reflection and elastic cotunneling are dressed
by local Andreev reflections. As a result ${\cal G}_{ab}^{\rm AP}$ can become negative
as the interface transparencies are increased, whereas ${\cal G}_{ab}^{\rm P}$ is always
negative.
We also carried out simulations of infinite planar interfaces
and found the same qualitative behavior without imposing
phase averaging by hand and without using the approximation
in which a single non local process is retained.

\section*{Acknowledgments}
One of the authors (D.F.) acknowledges stimulating discussions with
G. Deutscher and F. Sols.

\appendix

\section{Evaluation of integrals for infinite planar interfaces}
\label{app:change}
In this Appendix we give necessary technical details regarding
changes of variable for evaluating the current with infinite
planar interfaces.
The ``11'' component of the CAR current takes the form
\begin{eqnarray}
\label{eq:I11}
&&I^{\rm CAR}_{1,1} = 4 \pi^3 t_a^2 t_b^2 \frac{e}{h}
\left(\frac{2 m a_02}{\hbar2}\right)2 \int_{-e V_b}^{e V_a}
d \omega \int_{-D}^{D} d \xi_\parallel\\
&&\rho^{a,a}_{1,1}(\omega-e V_a,\xi_\parallel)
\rho^{b,b}_{2,2}(\omega-e V_b,\xi_\parallel)
\left| G^{A,\alpha,\beta}_{1,2}(\omega,\xi_\parallel) \right|^2
\nonumber
,
\end{eqnarray}
and the ``11'' component of the EC current takes the form
\begin{eqnarray}
\label{eq:IEC11}
&&I^{\rm EC}_{1,1} = 4 \pi^3 t_a^2 t_b^2 \frac{e}{h}
\left(\frac{2 m a_0^2}{\hbar^2}\right)^2 \int_{e V_b}^{e V_a}
d \omega \int_{-D}^{D} d \xi_\parallel\\\nonumber
&&\rho^{a,a}_{1,1}(\omega-e V_a,\xi_\parallel)
\rho^{b,b}_{1,1}(\omega-e V_b,\xi_\parallel)
\left| G^{A,\alpha,\beta}_{1,1}(\omega,\xi_\parallel) \right|^2
\end{eqnarray}
Similar expressions are obtained for the ``22'' currents and for the
currents in the spin-down sector.

The density of state $\rho^{a,a}_{1,1}(\omega-e V_a,\xi_\parallel)$
and $\rho^{b,b}_{1,1}(\omega-e V_b,\xi_\parallel)$
deduced from Eq.~\ref{eq:g-F-ip}
contain a square root singularity. For instance
\begin{equation}
\label{eq:rho-aa-11}
\rho^{a,a}_{1,1}(\omega-e V_a,\xi_\parallel)
=\frac{\sqrt{2 m a_0^2}}{\hbar}
\frac{\theta(\omega-e V_a-\xi_\parallel+h_a)}
{\sqrt{\omega-e V_a-\xi_\parallel+h_a}}
\end{equation}
After a change of variable $I^{\rm CAR}_{1,1}$ takes the form
\begin{eqnarray}
\label{eq:inter}I^{\rm CAR}_{1,1} &=&
8 \pi^3 t_a^2 t_b^2 \frac{e}{h}
\left(\frac{2 m a_0^2}{\hbar^2} \right)
\int_{x_{\rm min}}^{x_{\rm max}} dx
\int_0^{u_{\rm max}} \\ \nonumber
&&\frac{du}{\sqrt{2 x + u^2}}
\left| G^{\alpha,\beta}_{1,2}(\omega(x),\xi_\parallel) \right|^2
,
\end{eqnarray}
where explicit expressions of $x(\omega)$,
$x_{\rm min}$, $x_{\rm max}$ and $u_{\rm max}$ can be obtained in
each case. For instance if $-e V_b < e V_a$ we have
$x(\omega)=-\omega+e(V_a+V_b)/2-(h_a+h_b)/2$
and $u=\sqrt{\omega-e V_a - \xi_\parallel +h_a}$.
The integral (\ref{eq:inter}) is then evaluated by
making the changes of variable $v^2=2x$ and $(u,v)=\rho (\cos{\theta},
\sin{\theta})$, therefore absorbing the square root singularity
in a change of variable.


\begin{thebibliography}{99}
\bibitem{error} A mistake is present in Ref. \onlinecite{Melin-Feinberg-CAR},
in the last matrix of expression (12).
\end{thebibliography}
\end{document}